\documentclass[twocolumn]{aastex62}

\shorttitle{Wideband polarized radio emission from XTE J1810$-$197}
\shortauthors{Dai et al.}

\begin{document}

\title{Wideband polarized radio emission from the newly revived magnetar XTE J1810$-$197}

\correspondingauthor{Shi Dai}
\email{shi.dai@csiro.au}

\author[0000-0002-9618-2499]{Shi Dai}
\affiliation{CSIRO Astronomy and Space Science, Australia Telescope National Facility, Epping, NSW 1710, Australia}

\author[0000-0001-9208-0009]{Marcus E. Lower}
\affiliation{Centre for Astrophysics and Supercomputing, Swinburne University of Technology, PO Box 218, Hawthorn, VIC 3122, Australia}
\affiliation{CSIRO Astronomy and Space Science, Australia Telescope National Facility, Epping, NSW 1710, Australia}

\author[0000-0003-3294-3081]{Matthew Bailes}
\affiliation{Centre for Astrophysics and Supercomputing, Swinburne University of Technology, PO Box 218, Hawthorn, VIC 3122, Australia}
\affiliation{OzGrav: The ARC Centre of Excellence for Gravitational-wave Discovery, Hawthorn VIC 3122, Australia}

\author[0000-0002-1873-3718]{Fernando Camilo}
\affiliation{South African Radio Astronomy Observatory, Observatory 7925, South Africa}
%\affiliation{SKA South Africa, Pinelands 7405, South Africa}

\author{Jules P. Halpern}
\affiliation{Columbia Astrophysics Laboratory, Columbia University, New York, NY 10027, USA}

\author[0000-0002-7122-4963]{Simon Johnston}
\affiliation{CSIRO Astronomy and Space Science, Australia Telescope National Facility, Epping, NSW 1710, Australia}

\author[0000-0002-0893-4073]{Matthew Kerr}
\affiliation{Space Science Division, Naval Research Laboratory, Washington, DC 20375-5352, USA}

\author{John Reynolds}
\affiliation{CSIRO Astronomy and Space Science, Australia Telescope National Facility, Epping, NSW 1710, Australia}

\author{John Sarkissian}
\affiliation{CSIRO Astronomy and Space Science, Parkes Observatory, PO Box 276, Parkes NSW 2870, Australia}

\author{Paul Scholz}
\affiliation{National Research Council of Canada, Herzberg Astronomy and Astrophysics, Dominion Radio Astrophysical Observatory, Penticton, BC V2A 6J9, Canada}

%\submitjournal{ApJL}

\begin{abstract}

The anomalous X-ray pulsar XTE J1810$-$197 was the first magnetar found to emit pulsed radio emission. 
After spending almost a decade in a quiescent, radio-silent state, the magnetar was reported to have undergone a radio outburst in December, 2018.
We observed radio pulsations from XTE J1810$-$197 during this early phase of its radio revival 
using the Ultra-Wideband Low receiver system of the Parkes radio telescope, obtaining wideband (704\,MHz to 4032\,MHz) polarization pulse profiles, single pulses and flux density measurements. 
Dramatic changes in polarization and rapid variations of the position angle of linear polarization across the main pulse and in time have been observed. 
The pulse profile exhibits similar structures throughout our three observations (over a week time scale), displaying a small amount of profile evolution in terms of polarization and pulse width across the wideband. 
We measured a flat radio spectrum across the band with a positive spectral index, in addition to small levels of flux and spectral index variability across our observing span.
The observed wideband polarization properties are significantly different compared to those taken after the 2003 outburst, and therefore provide new information about the origin of radio emission.
\end{abstract}


\section{Introduction} \label{sec:intro}

Magnetars are a sub-class of pulsars with long rotation periods and high spin-down rates,
which regularly undergo X-ray outbursts where the energy released exceeds the 
spin-down luminosity~\citep[see e.g.,][for reviews]{kb17}.
The anomalous X-ray pulsar (AXP) XTE J1810$-$197, discovered in early 2003 following an X-ray outburst~\citep{ims+04}, was the first magnetar identified to be a transient pulsating radio source~\citep{crh+06}.
During the early phase of the previous radio outburst, extremely variable flux densities and pulse profiles with flat spectra were observed~\citep{crp+07,ccr+07,crj+07,ksj+07,ljk+08}. 
While different from those of normal radio pulsars, these properties are shared by all four radio magnetars identified so
far~\citep{crh+07,crj+08,lbb+10,kjl+11,sj13}, and are similar to those of the high magnetic field pulsar PSR~J1119$-$6127 following its magnetar-like X-ray bursts~\citep{djw+18}. 
Since the initial detection of its pulsed radio emission, XTE J1810$-$197 has been regularly monitored by multiple radio telescopes~\citep[e.g.,][]{crh+16}.
After rapidly decreasing over the first 10 months, the averaged radio flux density remained steady for the next 22 months before fading and eventually disappearing in late 2008~\citep{crh+16}.

For the following 10 years, XTE J1810$-$197 remained in a quiescent/low-activity state 
in both radio and X-ray bands~\citep{pme+19}, until intense radio emission was reported to have restarted 
sometime between 2018 October 26 and December 8 \citep{lls+18}. Follow-up observations have been carried 
out in radio and X-ray bands by several telescopes, where strong pulsations in both bands 
have been detected~\citep[e.g.,][]{dek+18,lbj+18,ghg+18,gha+19}.
Early observations of the outburst and radio revival are of great importance 
for understanding the origin of magnetar radio emission and to study their behaviour post-outburst.  
Here we present observations (on 2018 December 11, 15 and 18) of XTE J1810$-$197, three days after the first report of its revival, using the Ultra-Wideband Low (UWL) receiver system recently installed at the Parkes 64 meter radio telescope~\citep{hob+19}. We 
show wideband polarization pulse profiles, single pulses and flux density measurements 
of the magnetar during the early phase of its revival, and compare these initial results 
with previous observations. 

\section{Observations and data reduction}
\label{sec:observations}

XTE J1810$-$197 was observed on 2018 December 11, 15 and 18 at Parkes using the UWL receiver, which provides radio frequency coverage from $704$\,MHz to 4032\,MHz. 
Further observations using the UWL are ongoing and will be the subject of future work.
The December 11 observation was performed using the transient search mode where data is recorded with 8-bit sampling every 128\,$\mu$s 
in each of the 1\,MHz wide frequency channels (3328 channels across the whole band). 
Observations on December 15 and 18 used the pulsar fold mode where data is folded modulo the pulse period with 1024 phase bins in each of the 1\,MHz channels, integrated for 30\,s and written to disc. 
For both modes, data were coherently de-dispersed at a dispersion measure (DM) of $178 \pm 5$\,pc\,cm$^{-3}$~\citep{crh+06} with full Stokes information recorded. 
Total integration times are 819, 1770 and 631 seconds for observations on December 11, 15 and 18, respectively. 

At the current stage, a critical sampling filter bank has been used to produce 26 sub-bands. We removed 5\,MHz of the bandpass at each edge of the 26 sub-bands to mitigate aliasing. 
We manually excised data affected by narrow-band and impulsive radio-frequency interference (RFI) for each sub-integration. 
To measure the differential gains between the signal paths of the two voltage probes, we observed a pulsed noise signal injected into the signal path prior to the first-stage low-noise amplifiers before each observation. 
The noise signal also provides a reference brightness for each observation. 
To correct for the absolute gain of the system, we use
observations of the radio galaxy 3C 218 (Hydra A); using on- and off-source pointings to measure the apparent brightness of the noise diode as a function of radio frequency.
Polarimetric responses of the UWL are derived from observations of PSR J0437$-$4715~\citep{jlh+93} covering a wide range of parallactic angles~\citep{v04}, taken during the commissioning of UWL in November 2018. 
The Stokes parameters 
are in accordance with the astronomical conventions described by \citet{vmj+10}.
The linear polarization and the position angle of linear polarization were calculated 
following \citet{dhm+15}. Search-mode data is 
folded using the {\sc dspsr}~\citep{vb11} software package. All data reduction and 
calibration used the {\sc psrchive}~\citep{hvm04} software package.

\section{Results}
\label{sec:result}

\begin{figure*}
\centering
\includegraphics[width=18cm]{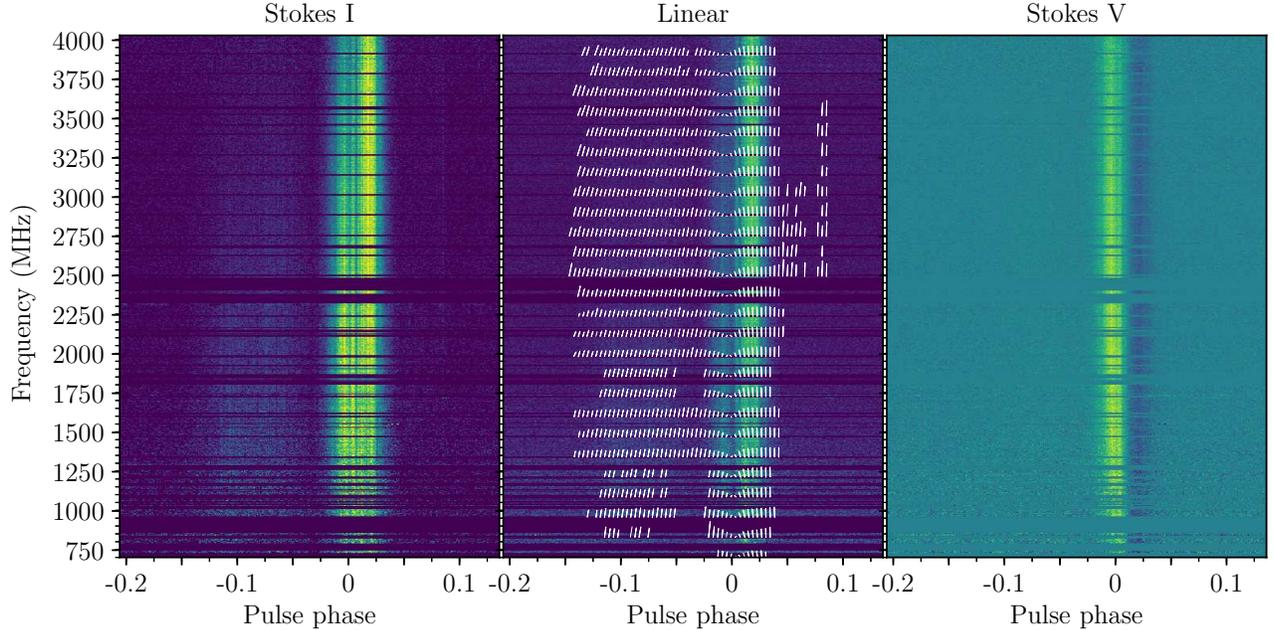}
\caption{Average polarization pulse spectra for XTE J1810$-$197 taken 
on 2018 December 11. From left to right, we show Stokes I, linear polarization 
and Stokes V, normalised by their peak values. Gaps represent zero-weighted 
channels that were consistently or 
strongly contaminated by RFI. The PA of the linear polarization (over each 
128\,MHz subband) is shown as white solid lines. The length of lines represent the fraction of linear polarization. 
}
\label{heatmap}
\end{figure*}

\begin{figure*}
\begin{center}
%\begin{tabular}{c}
%\includegraphics[width=16cm]{3392.ps}\\
%\includegraphics[width=16cm]{2624.ps}\\  
%\includegraphics[width=16cm]{1600.ps}  
%\end{tabular}
\includegraphics[width=18cm]{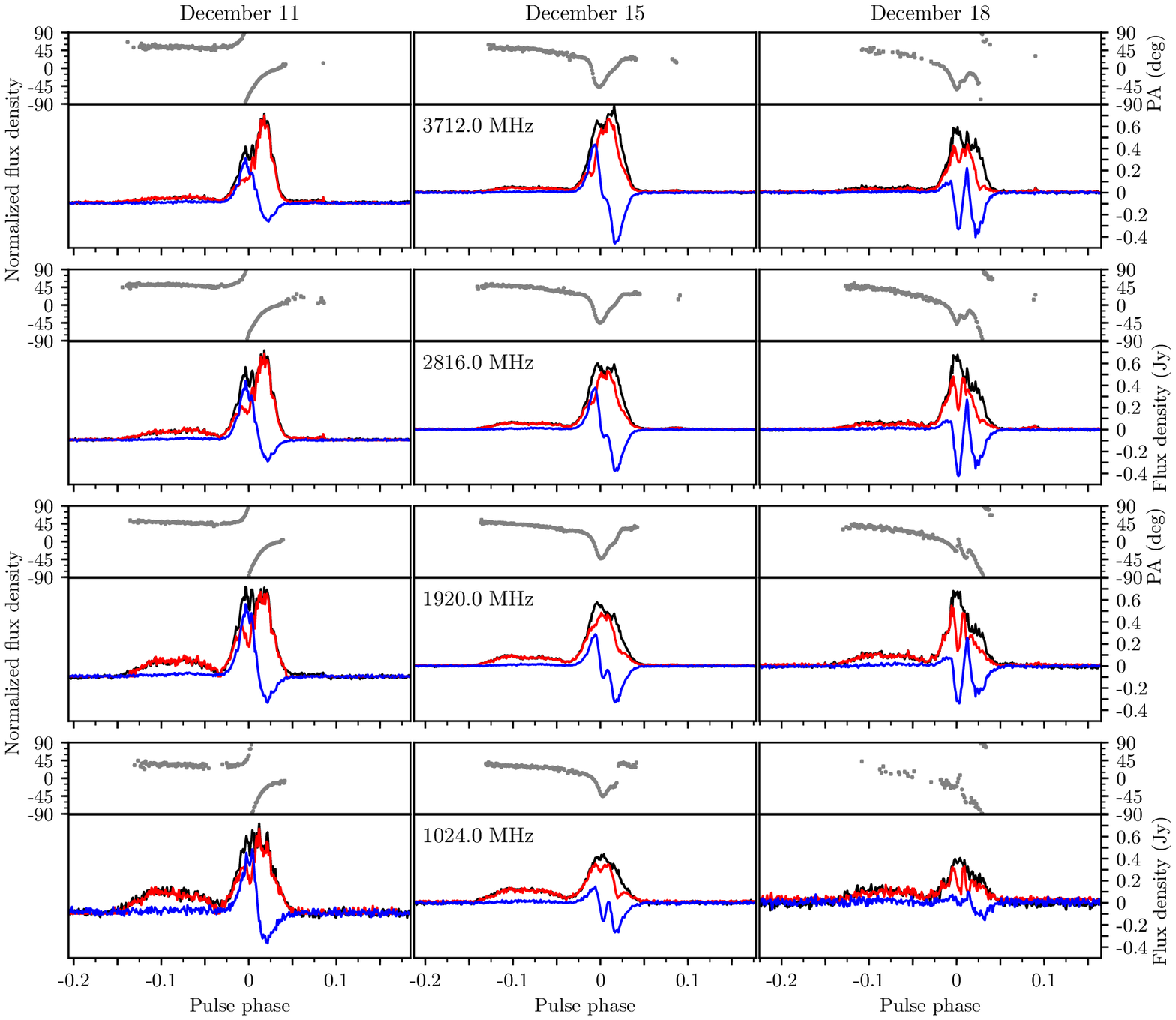}
\end{center}
\caption{Averaged polarization pulse profiles centred at 3712, 2816, 1920 and 1024\,MHz (each with a bandwidth of 128\,MHz) at three epochs. 
In each panel, the black line shows total intensity, red linear polarization, and blue circular polarization. 
The PA of the linear polarization is shown in the top panels.
Search-mode observation on December 11 is not flux calibrated, and therefore we show the normalized flux density.
}
\label{prof}
\end{figure*}

%\subsection{Polarization pulse profile}
%\label{sec:prof}

In Fig.~\ref{heatmap}, we show the total intensity, as well as 
the linearly and circularly polarized emission and the position angle (PA) 
of linear polarization, as a function of pulse phase and frequency for data 
taken on December 11. 
Across a continuous frequency coverage from 704\,MHz to 4032\,MHz, 
pulse profiles show remarkably small frequency evolution. Apart from 
the leading component and leading edge of the main pulse becoming 
shallower at higher frequencies and the trailing component disappearing 
at lower frequencies, the structure, polarization and PA remain 
similar across the band. 
We obtain a rotation measure (RM) of $74.44\pm0.16$~\,rad\,m$^{-2}$, which 
agrees with previous measurements~\citep{crj+07}. All the PA 
have been corrected for the measured RM, referred to infinite frequency.

To better present details of polarization pulse profiles, in Fig.~\ref{prof}, 
we show the average polarization pulse profile of our three observations 
at four different frequencies. The observed pulse 
profiles show three components, a wide and shallow leading component, a strong, 
spiky main pulse and a weak trailing component. The main pulse is similar to 
that observed after the 2003 outburst, which was found to be dominated 
by strong, narrow single pulses~\citep{ccr+07,ksj+07}. All three pulse 
components have high linear polarization across the entire band, consistent 
with previous observations. The leading and trailing components are 
almost 100\% linearly polarized. 
%

%\subsection{Single pulses}
%\label{sec:sp}

\begin{figure*}
\centering
\includegraphics[width=\linewidth]{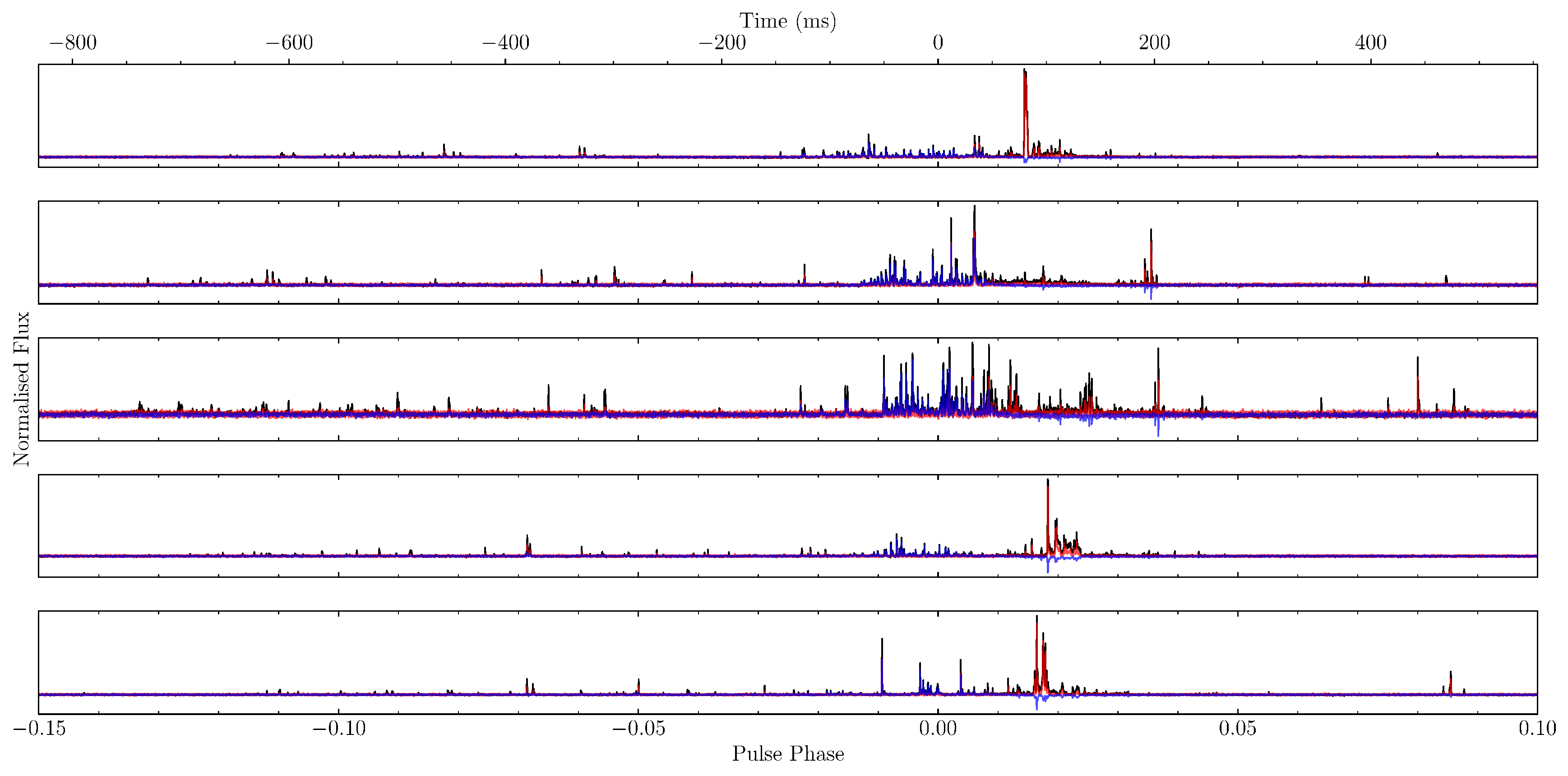}
\caption{A selection of high time resolution polarimetric single pulse profiles seen on December 11th, 2018. The integrated profiles show the total intensity in black, linear polarization in red and circular polarization in blue.}
\label{single_pulses}
\end{figure*}

Strong single pulses from XTE J1810$-$197 were recorded during the search-mode observation on December 11. Individual pulses are found to be comprised of narrow spiky emission components clustered around the phase of the main pulse and scattered throughout the leading and tail components. 
Similar to multi-frequency analyses during the 2003 outburst~\citep{ssw+09}, the amplitudes and number of emission spikes are seen to be highly variable from pulse-to-pulse.  
A selection of high time resolution, single pulse polarization profiles are presented in Fig.~\ref{single_pulses}. The single emission spikes display remarkable millisecond-width temporal structure and polarization evolution across the pulse cycle. 
Spikes within the leading, latter part of the main pulse, and tail components are almost completely linearly polarized, whereas those within the leading half of the main pulse show a large degree of circular polarization.
The single pulse fluxes remain relatively flat across the UWL band, with little evidence of frequency evolution with phase.
We note that single pulses were not recorded for the December 15 and 18 observations. Given the pulse-to-pulse variability seen in this single epoch, the single pulses presented here may not be representative of those in other observations.

An improved estimate of the magnetar's DM is obtained by generating times of arrival (ToAs) from the high resolution single-pulse archives 
in Fig.~\ref{single_pulses}. This is achieved by splitting the archives 
into 64 sub-bands and using the integrated single-pulse profiles as 
templates. Any residual dispersion in the pulse would result in an 
increasing delay in the ToAs generated from the high to low frequency 
sub-bands, which we correct for by fitting the DM using the reduced 
$\chi^{2}$ method implemented in {\sc tempo2}~\citep{hem06}. 
From this method we obtain a mean DM of $178.51 \pm 0.08$\,pc\,cm$^{-3}$ where the uncertainty is the variance between DM measurements for each pulse. 
This result represents an improvement over the previous measurement 
reported in \citet{crh+06} by two orders of magnitude.
We also obtain a spin-frequency measurement of $\nu = 0.180463 \pm 0.000007$\,Hz from this observation (MJD 58463.963).

%\subsection{Flux density and spectrum}
%\label{sec:spec}

\begin{table}[b!]
\begin{center}
\caption{UWL measurements of period-averaged flux densities.}
\label{flux}
\begin{tabular}{lcc}
\hline
\hline
Flux Density & December 15 & December 18 \\
\hline
(mJy) & & \\
\hline
$S_{768}$  & $18.8 \pm 0.3$  & $14.2 \pm 0.4$ \\
$S_{1024}$ & $22.6 \pm 0.2$  & $20.3 \pm 0.2$ \\
$S_{1280}$ & $25.2 \pm 0.2$  & $23.9 \pm 0.2$ \\
$S_{1536}$ & $26.8 \pm 0.1$  & $27.1 \pm 0.1$ \\
$S_{1792}$ & $28.9 \pm 0.1$  & $29.6 \pm 0.1$ \\
$S_{2048}$ & $29.3 \pm 0.1$  & $30.7 \pm 0.1$ \\
$S_{2304}$ & $30.05 \pm 0.08$& $30.9 \pm 0.1$ \\
$S_{2560}$ & $31.11 \pm 0.05$& $30.31 \pm 0.09$ \\
$S_{2816}$ & $31.64 \pm 0.05$& $30.36 \pm 0.09$ \\
$S_{3072}$ & $33.18 \pm 0.07$& $28.91 \pm 0.09$ \\
$S_{3328}$ & $34.2 \pm 0.1$  & $29.7 \pm 0.1$ \\
$S_{3584}$ & $36.4 \pm 0.1$  & $28.3 \pm 0.1$ \\
$S_{3840}$ & $35.7 \pm 0.2$  & $29.9 \pm 0.2$ \\
\hline
\end{tabular}
\end{center}
\end{table}

To measure the flux density, we formed noise-free standard templates for each observation using the {\sc psrchive} program {\sc paas} after integrating the data over the observing band and observation duration.
The {\sc psrchive} program {\sc psrflux} was used to measure the flux density for each observation, which cross-correlates the observed profile with the standard template to obtain the scaling factor and then the averaged flux density. 
The uncertainty of flux density was estimated using the standard deviation of the baseline fluctuations.
As described in Section~\ref{sec:observations}, for fold-mode observations, the absolute gain of the system was calibrated using observations of Hydra A. Flux density measurements of the fold-mode observations are listed in Table~\ref{flux}. 
For current UWL search-mode observations, the absolute gain can not be calibrated. We estimate the search-mode flux density by comparing the 
baseline root-mean-square (RMS) with that of the calibrated 
fold-mode observation taken on December 18, which was less affected by RFI.

\begin{figure}
\begin{center}
\vspace*{-1.5cm}
\includegraphics[width=10cm]{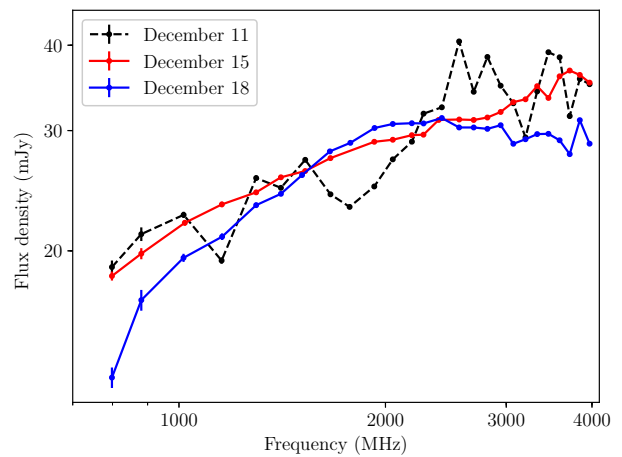}
\end{center}
\vspace*{-0.9cm}
\caption{Flux density as a function of frequency. For the search-mode
observation taken on 2018 December 11 (black points with dashed lines), 
flux densities were calibrated as described in Section~\ref{sec:result}. Red and blue points with solid connecting lines represent the flux density measured during fold-mode observations on December 15 and 18.}
\label{spec}
\end{figure}

With the UWL, we were able to simultaneously measure flux densities across 
a wideband and mitigate the effect of large flux density variations.
In Fig.~\ref{spec}, we show flux density measurements of our three 
observations. Large variations in search-mode observation flux densities 
are likely to be introduced by low level RFI in the baseline. Although 
the spectral shape shows significant variations, the spectra generally follow 
a power-law with positive spectral indices. Fitting a power law spectrum 
of the form $S_{\nu}\approx\nu^{\alpha}$ to measurements on December 15, 
we measure a spectral index of $\alpha\approx+0.3$.

\section{Discussion}

Our observations captured the wideband polarization properties of the magnetar 
during the early stages of its latest outburst. We observed several differences 
in the pulse profile properties when compared to those observed after 
the 2003 outburst~\citep{ccr+07,crj+07,ksj+07}. Firstly, the overall 
structure of the pulse profile did not show dramatic changes during 
our observations over a week period, with the leading and trailing 
components remaining stable.  
Secondly, we observed strong circular polarization in the main pulse and 
no evidence of an interpulse. However, the highly linearly polarized leading 
component with flat and stable PA swing is reminiscent of the previously 
observed interpulse.
Finally, unlike the relatively stable polarization properties 
of the main pulse reported by \citet{crj+07} and \citet{ksj+07}, we observed 
significant changes in both the linear and circular polarization across the 
main pulse. The PA swing across the main pulse changed in both shape 
and absolute value over a week period, and show complex structures 
which deviate from a simple rotating vector model~\citep[RVM,][]{rc69}.
Interestingly, the PA of the leading and trailing components show 
almost no variation in time at all. 

When compared to observations after the previous outburst, the difference in polarization properties we observe might indicate structural changes in the magnetosphere 
triggered by the latest outburst, with the possibility that we are observing 
radio emission from a new active region.
Changes in the PA swing were also observed in PSR J1622$-$4950 after 
its recent radio activation in 2017,
which can generally be modelled by a RVM with variations explained by different 
emission locations within the magnetosphere~\citep{lbb+12,css+18}. 
However, the observed PA swing of XTE J1810$-$197 shows complex structures 
and does not follow the RVM model obtained during the previous outburst~\citep{crj+07}. 
In addition, by comparing the radio and X-ray profiles, \citet{gha+19} show that 
the macroscopic geometry of the magnetic field associated with the emission 
did not change between outbursts.

Alternatively, it has been suggested that magnetospheres of magnetars can become twisted by sudden crustal motion, where the untwisting of magnetic field lines 
dissipates magnetic energy and produces electromagnetic radiation~\citep[e.g.,][]{bel09}. 
In this scenario, radio emission may originate on the bundle of 
closed field lines, and large variations in polarization properties
can be explained by the instabilities in the twisted magnetosphere. While the polarization and PA were observed to change on a time scale of days, we note the overall shape of total intensity remained relatively similar, including the profile width and separation between different components. 

Furthermore, we observe little frequency evolution in the pulse profile and 
polarization properties over our wide band. \citet{crj+07} compared the pulse 
profile during the previous outburst at 1.4, 3.2 and 8.4\,GHz, showing the double-peaked profile became 
narrower at higher frequencies, and the ratio of the leading to trailing component 
becomes larger. Here we also observe the ratio of the leading to main component becomes 
larger, but find no evidence of profile narrowing at higher frequencies.
The lack of profile and polarization evolution over such a wide band 
puts tight constraints on current models of the origin of radio emission 
from magnetars. For the twisted magnetosphere model, twists of magnetic fields are 
expected to be larger as one goes further along a given field line, which would result in observable differences in the polarization profile as a function of frequency.

For XTE J1810$-$197, significant variation of PA swing and absolute 
PA values have only been observed for the interpulse after the 2003 
outburst~\citep{ksj+07}, which is argued to be evidence of 
magnetospheric propagation effects in a non-dipolar magnetic field 
configuration. Such propagation effects might also explain the origin 
of circular polarization not observed before. If the PA swing variations 
are caused by magnetospheric propagation effects, the three adjacent 
pulse components we observe must originate from very different locations
in the magnetosphere and the dramatic variation we observed implies 
very strong propagation effects at the early stage of the outburst.

A similar example of dramatic changes in polarization within a short 
period of time was observed in the high magnetic field pulsar J1119$-$6127 
after a magnetar-like X-ray burst. The dramatic profile and PA variations of 
PSR J1119$-$6129 was accompanied by peculiar spin-frequency evolution and 
large fluctuation of flux density, which have been suggested to be linked 
with either reconfiguration of magnetic field or strong particle
winds~\citep{djw+18}. Similar changes in the magnetosphere could also 
be triggered by the outburst of XTE J1810$-$197, and produced the 
observed polarization variation. 

The radio spectrum of XTE J1810$-$197 after the 2003 outburst was studied 
by \citet{crp+07,ljk+08}, where a spectral index of $-0.5\lesssim\alpha\lesssim0$ 
over the range of 1.4 to 144\,GHz was observed. We observed a harder 
spectrum than what was measured after the 2003 outburst.
The spectrum shape showed changes with time over the ten day observing span, 
but we find no evidence of the dramatic variations in 
$\alpha$ that have previously been observed in XTE J1810$-$197~\citep[e.g.,][]{ljk+08} 
and other radio magnetars~\citep{crj+08,lbb+10,ppe+15}. On December 18, we 
observe evidence of a spectral turnover at around 2\,GHz.

At 1.4\,GHz, we measured an average flux density of 
$\sim20$\,mJy at all three epochs, which is about a factor of two higher
than those measured at the first detection of radio pulsations from 
XTE J1810$-$197~\citep{crh+06}. Observations beginning three years 
after the 2003 outburst showed that the flux density decreased 
by a factor of 20 during 10 months in 2006.
Similarly, measurements of the magnetar PSR J1622$-$4950
soon after its radio revival in early 2017 reached a maximum averaged 
flux density of tens of mJy at 1.4 to 3\,GHz~\citep{css+18}.
These suggest that the wide band observations
presented here might be during the most radio-active period of the current outburst.

\section{Summary}
\label{sec:summary}

After 10 years in a radio silent state, pulsed radio emission from the magnetar XTE J1810$-$197
was found to have reactivated in early December 2018. We conducted follow-up 
observations using the UWL receiver of the Parkes 
telescope on 2018 December 11, 15 and 18. Ultra-wideband polarization pulse 
profiles, single pulses and flux density measurements during the early 
phase of the outburst are presented in this paper. While pulse profiles of 
the total intensity show similar structures during the three epochs, the polarization 
properties show significant variability. 
We also find the PA swing varied dramatically across the main pulse and in time, however it remained stable across the leading and trailing pulse components. 
Despite dramatic variations in time, very little frequency evolution of the polarization have been observed across the wideband. 
Compared with measurements after the previous outburst, we find the flux density of XTE J1810$-$197 to be at least a factor of two higher,which could be due to our observations capturing an earlier stage of the radio outburst.
Radio spectra across the wideband generally follow a power-law with positive spectral indices, and show small levels of variation across the observation epochs. 
The wideband polarization properties are quite different 
to what was seen during the previous outburst and showed 
dramatic variations in time, which might indicate rapid changes in 
the magnetosphere soon after the outburst.
Future multi-wavelength, long term monitoring of this pulsar is required to complete the full picture of the magnetar outburst behaviour.

\acknowledgments

We thank the Parkes team for their great efforts during the installation 
and commissioning of the UWL receiver system. We thank Maxim Lyutikov 
for useful discussions. 
The Parkes radio telescope is part of the Australia Telescope National 
Facility which is funded by the Commonwealth of Australia for operation 
as a National Facility managed by CSIRO. Part of this 
work made use of the OzSTAR national HPC facility. Work at NRL is 
supported by NASA. We also acknowledge use of the Astronomer's 
Telegram and the NASA Astrophysics Data Service. 

%\bibliography{ms}


\end{document}